%% file: vrst_main.tex
\documentclass[sigconf]{acmart}
\renewcommand\footnotetextcopyrightpermission[1]{} 
\AtBeginDocument{%
  }

\usepackage{orcidlink}
\usepackage{graphicx}
\begin{document}

\title{Motion-Based User Identification across XR and Metaverse Applications by Deep Classification and Similarity Learning}

\renewcommand{\shorttitle}{Motion-Based User Identification across XR and Metaverse Applications by Deep Classification and Similarity Learning}
\author{Lukas Schach \orcidlink{0009-0002-1876-1994} }
\orcid{0009-0002-1876-1994}
\email{lukas.schach@uni-wuerzburg.de}
\affiliation{%
  \institution{University of Würzburg}
  \country{Germany}}

\author{Christian Rack \orcidlink{0000-0002-0022-0711} }
\email{christian.rack@uni-wuerzburg.de}
\orcid{0000-0002-0022-0711}
\affiliation{%
  \institution{University of Würzburg}
  \country{Germany}}

\author{Ryan P. McMahan \orcidlink{0000-0001-9357-9696} }
\orcid{0000-0001-9357-9696}
\email{rpm@vt.edu}
\affiliation{%
  \institution{Virginia Tech Blacksburg}
  \country{United States}}

\author{Marc Erich Latoschik \orcidlink{0000-0002-9340-9600}}
\orcid{0000-0002-9340-9600}
\email{marc.latoschik@uni-wuerzburg.de}
\affiliation{%
  \institution{University of Würzburg}
  \country{Germany}}
  
\renewcommand{\shortauthors}{Schach et al.}

\begin{abstract}
This paper examines the generalization capacity of two state-of-the-art classification and similarity learning models in reliably identifying users based on their motions in various Extended Reality (XR) applications. 
We developed a novel dataset containing a wide range of motion data from 49 users in five different XR applications: four XR games with distinct tasks and action patterns, and an additional social XR application with no predefined task sets. The dataset is used to evaluate the performance and, in particular, the generalization capacity of the two models across applications. Our results indicate that while the models can accurately identify individuals within the same application, their ability to identify users across different XR applications remains limited. Overall, our results provide insight into current models generalization capabilities and suitability as biometric methods for user verification and identification. The results also serve as a much-needed risk assessment of hazardous and unwanted user identification in XR and Metaverse applications. Our cross-application XR motion dataset and code are made available to the public to encourage similar research on the generalization of motion-based user identification in typical Metaverse application use cases.

\end{abstract}



\begin{teaserfigure}
  \includegraphics[width=\textwidth]{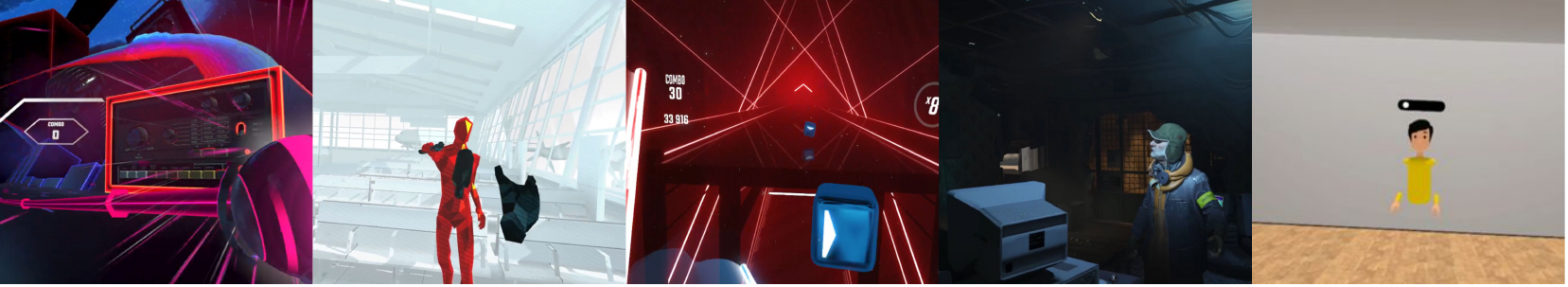}
  \caption{Screenshots from all five VR applications, arranged from left to right in the order participants played them: Synth Riders, Superhot VR, Beat Saber, Half-Life: Alyx, and a Social VR scenario.}
  \Description{}
  \label{fig:all_vr_applications}
  \label{fig:teaser}
\end{teaserfigure}


\maketitle
\pagestyle{plain} 
\input{sections/01-introduction}

\input{sections/02-related-work}
\input{sections/03-machinelearning}

\input{sections/04-datasets}

\input{sections/05-evaluation}

\input{sections/06-results}

\input{sections/07-discussion}

\bibliographystyle{ACM-Reference-Format}
\bibliography{new_main_bib}

\end{document}

%% file: sections/01-introduction.tex
\section{Introduction}

Users of Virtual Reality (VR), Augmented Reality (AR), and Mixed Reality (MR), or eXtended Reality (XR) for short, inherently share their motion data with the system, an inevitable requirement for delivering an immersive experience.
Recent research has demonstrated that this motion data---such as tracking information from Head-Mounted Displays (HMDs) and hand controllers---contains identifiable patterns, enabling modern machine learning models to identify users solely based on their movements (e.g.,~\cite{rogersApproachUserIdentification2015,millerWithinSystemCrossSystemBehaviorBased2020,rackWhoAlyx2023,nairUniqueIdentification500002023,baldoniMovementTrafficbased2025}). 
These capabilities open up two prominent use cases: motion-based user identification can serve as a useful tool for biometric user verification to grant system and resource access. 
However, it also raises a potential risk factor of hazardous and unwanted user identification in use cases where anonymity is wanted.
This has sparked significant concerns regarding privacy in a Metaverse~\cite{nairEffectDataDegradation2024, nairTruthMotionUnprecedented2024, millerPersonalIdentifiabilityUser2020}, where (personal) motion data is also shared across machines in distributed settings of a social XR.    

Preliminary work already achieved a remarkable high accuracy in user identification \cite{rackVersatileUser2024, rogersApproachUserIdentification2015, millerWithinSystemCrossSystemBehaviorBased2020, rackWhoAlyx2023, nairUniqueIdentification500002023}. However, these results were achieved on motion patterns generated within the same applications. But how critical are the potential threats of unwanted user identification today, taking into account the various application types envisioned for a future Metaverse? Concretely, is it possible to identify individuals across different XR applications? This would have profound implications for privacy.
For example, numerous players of the Beat Saber VR game upload replays of their sessions every day, including complete tracking data, to public scoreboards like Beat Leader\footnote{\url{https://beatleader.xyz}} to compare and track their performances.
Malicious actors could leverage motion-based identification models to build motion profiles of these players, making it impossible for them to remain anonymous in other parts of the Metaverse.

Addressing the threat to privacy posed by state-of-the-art data-driven user identification across different VR applications requires appropriate datasets that capture the essential motion variances. However, openly available datasets did not provide movement data for a sufficient number of users across different applications that offer the required variance in movement patterns (see Table \ref{tab:dataset}). Hence, we created a new dataset from users who played five different VR applications specifically curated to meet these requirements. The selected VR applications vary significantly in the types of movements they require, ranging from rhythm games like Beat Saber and Synth Riders, which require precise, specific actions, to more open-ended experiences like Half-Life: Alyx and SuperHot VR, where movements are less constrained and more varied. 
Additionally, we collected data from diadic multi-modal conversations between participants in a social VR scenario. This setting captures more natural and unscripted movements, providing a contrast to more structured gameplay, and it additionally represents a potential prominent use case of a future Metaverse. 


\hspace{1em}

\noindent\textbf{Contributions}

\noindent Altogether, this article contains the following contributions:

\begin{itemize}
    \item Evaluation of two state-of-the-art models, i.e., similarity learning and classification learning models, in their ability to identify users across different VR applications based on their movements.
    \item Creation of a new dataset with 49 users and over 60 hours of motion data, collected across five VR applications: Beat Saber, Half-Life: Alyx, Synth Riders, Superhot VR, and a multi-modal conversation in a social VR environment.  
\end{itemize}


Our results show that the two prominent state-of-the-art models yet
exhibit significant limitations in their generalization capacity for user identification across VR applications. Hence, as of now, we consider the threat to privacy posed by such methods moderate at best. However, our across-application dataset and potentially upcoming comparable developments open up future research and development opportunities for such capacities, while simultaneously providing a means to assess and monitor the associated threat vector to privacy. We will make the dataset and the entire code for conducting the evaluation available on GitHub upon publication.

%% file: sections/02-related-work.tex
\begin{table*}[tb]
    
\centering
\caption{This table summarizes published motion datasets, showing the number of users, applications, and how many each user used. Ranges like 1–2 or 1–4 indicate user-specific variation in app usage.}
\label{tab:dataset}
\begin{tabular}{l|l|l|l|l}
\hline
ref &  Number of   & Number of     &   Number of             & Applications \\ 
    & Users        & Applications  & Applications per User   &   \\
\hline
Rack et al.\cite{rackWhoAlyx2023} & 71 &  1 & 1 & Half-Life: Alyx  \\
Moore et al. \cite{moore2023} & 108 & 1 &  1 & Assembly Tasks \\
Liebers et al. \cite{liebersIdentifyingUsersTheir2024} & 16 &  1 & 1 & Bowling and Archery  \\
Liebers et al. \cite{liebersExploringStabilityBehavioral2023} & 15 & 1 &  1 & Beat Saber \\
Liebers et al. \cite{liebersUnderstandingUserIdentification2021a} & 16 & 1 &  1 & Interacting with various interfaces  \\
Miller et al. \cite{millerUsingSiameseNeural2021} & 41 & 1 &  1 & throwing a ball \\
Nair et al. \cite{nair2023} & 100,000 &  2 &  1 & Beat Saber and Tilt Brush  \\
Rack et al. \cite{RackMotionPasswords2024} & 48 &   1 &  1 & Motion Password \\
Wen et al. \cite{wenVRNetRealworld2024} & 16 &  8 &  1-2 & Beat Saber, Cartoon Network, ...   \\
Thiel et al. \cite{thiel2022} & 20 &  7 &  1-4 & Beat Saber, Clash of Chesfs,  ... \\
Baldoni et al. \cite{baldoniMovementTrafficbased2025} & 60 &  4 &  2 & Beat Saber,  Cooking Simulator,  ... \\
\hline
Our Dataset & 49 &  5 & 5 &  Half-Life: Alyx, Superhot VR  \\
& & &   & Beat Saber, Synth Rider, Social VR \\
\hline
\end{tabular}
\end{table*}

\section{Related Work}
Motion-based user identification extends biometric analysis to encompass human motions to create a representation of an individual's unique physiological and/or behavioral characteristics.
In this context, we focus on typical XR systems that track the head and at least one hand.
Following the terminology defined by Jain et al.~\cite{jainIntroductionBiometrics2011}, biometric user identification systems can be used for either identification or verification tasks. 
Verification involves confirming or denying a user's claimed identity, such as when logging into an account. 
Identification involves determining a user's identity from a set of known identities, which is typically relevant for applications such as content personalization or advertising.
Given our focus on the identification setting, we present related work on motion-based user identification in XR and summarize publicly available datasets.

\subsection{Motion-Based User Identification in XR}
To guide through the landscape of current motion-based user identification approaches and distinguish between non-pretrainable and pretrainable methods.
Non-pretrainable methods are represented by classification learning models that are directly trained on data from the users they should identify at a later point.
Roger et al.~\cite{rogersApproachUserIdentification2015} were the first to use motion data to identify 20 individuals. 
Miller et al.~\cite{millerWithinSystemCrossSystemBehaviorBased2020} showed that this approach can be scaled to work with 511 users. 
Rack et al.~\cite{rackWhoAlyx2023} showed that deep learning methods can also be used for identification, and Nair et al.~\cite{nairUniqueIdentification500002023} demonstrated that it is possible to identify up to 50,000 users with 94.33\% accuracy using 100 seconds of motion data.
Baldoni et al.~\cite{baldoniMovementTrafficbased2025} have shown that users can be identified by combining motion data with traffic data.
They trained a separate model for each application and demonstrated user identification within a single application with accuracies exceeding 90\% in some cases.
They also briefly examined whether users could be identified across two applications and reported an accuracy of around 30\% in this scenario.
This aspect was not further explored in their study.
A severe barrier to the real-world applicability of such non-pretrainable methods is that onboarding is expensive since models require retraining for each new user, which takes significant time and resources.
Classification learning models also cannot indicate an `unknown' user, as they will always predict a user from the training dataset.

In contrast, pretrainable methods, such as feature-distance and similarity learning, can immediately be used to onboard and identify users without needing expensive retraining.
These methods work by producing some distance metric that will be small for motion sequences of the same users and large for different (or unknown) users.
Li et al.~\cite{liWhoseMoveIt2016} demonstrated that feature-distance methods could verify individuals with 95.57\% accuracy based on head movements while listening to music. 
Miller et al.~\cite{millerUsingSiameseNeural2021} showed that similarity learning could potentially identify individuals across different types of VR systems.
Rack et al.~\cite{rackVersatileUser2024} showed that similarity learning can indeed be pre-trained and immediately onboard new users, positioning motion-based user identification as a viable solution for real-world applications.
Subsequently, Rack et al.~\cite{RackMotionPasswords2024} showed that similarity learning shows comparative performance to a feature-distance model on uniform ball-throwing motion sequences but significantly outperforms complex handwritten in-air signatures.
In our paper, we evaluate both non-pretrainable and pretrainable approaches using a state-of-the-art transformer architecture.  

\subsection{Existing VR Motion Tracking Datasets}
We list previously published VR motion tracking datasets in table~\ref{tab:dataset} and compare them based on the number of users and VR applications.
Thus, we differentiate between datasets that feature either specific or nonspecific movements.
This distinction is also reflected in the number of possible activities that can be carried out in the different datasets.
VR applications with specific movements typically allow for a limited range of activities.
In contrast, VR applications with unspecific movements tend to support a broader variety of possible activities.
However, the exact number of activities is often only an estimate, as activities may overlap and are sometimes difficult to distinguish clearly.

In datasets with nonspecific movements, participants are not required to follow predefined actions. The tasks are only loosely defined in terms of timing and speed, resulting in a wide variety of activities that users can perform within the VR application. 
For example, Rack et al.~\cite{rackWhoAlyx2023} create a dataset from 71 users playing Half-Life: Alyx. Moore et al.~\cite{moore2023} created a dataset that involved 45 users performing assembly tasks, which they recently expanded to include 108 total participants~\cite{moore2024FAST}.
Wen et al.~\cite{wenVRNetRealworld2024} introduced the VR.net dataset, which included 16 users playing various VR games; however, this dataset is no longer publicly available, and only 5 of these users participated in the same two VR applications.
Thiel et al.~\cite{thiel2022} presented a dataset of 20 participants who played seven VR applications, including Beat Saber, Half-Life: Alyx, Pistol Whip, and Clash of Chefs. Participants were free to choose which applications to play, resulting in each application being played by 2 to 14 users. Only one participant played four different VR applications, while the others played three or fewer, which were not always the same.

On the other hand, the VR application often dictates specific movements or relies on an external source, typically at a predetermined time.
This usually limits the range of possible activities in the corresponding VR applications.
Several datasets have been created to capture these movements.
Liebers et al.~\cite{liebersIdentifyingUsersTheir2024,liebersExploringStabilityBehavioral2023, liebersUnderstandingUserIdentification2021a} developed several distinct datasets, including one focusing on 16 users performing bowling and archery tasks, another featuring 15 users playing Beat Saber, and a third capturing 16 users interacting with various interfaces, such as buttons and sliders. 
Miller et al.~\cite{millerUsingSiameseNeural2021} contributed a dataset involving 41 users throwing a ball using different devices. 
Nair et al.~\cite{nair2023} created the BOXRR-23 dataset, consisting of 100,000 users playing Beat Saber and Tilt Brush.
Baldoni et al.~\cite{baldoniMovementTrafficbased2025} created a dataset with 60 users, where one group of 30 played Beat Saber and Cooking Simulator. The other 30 users played Medal of Honor: Above and Beyond and Forklift Simulator.
Lastly, Rack et al.~\cite{RackMotionPasswords2024} provided a dataset where 48 users each entered 80 different writing sequences.

As seen in table~\ref{tab:dataset}, most datasets are restricted to single VR applications. 
Only three datasets~\cite{wenVRNetRealworld2024, thiel2022, baldoniMovementTrafficbased2025} capture users interacting with multiple VR applications. 
However, only a small number of users were recorded using the same set of VR applications, which prohibits a comprehensive evaluation of cross-application identification.
We contribute to the landscape of existing datasets by publishing our own dataset consisting of 49 users using five different VR applications.


%% file: sections/03-machinelearning.tex
\section{Concept: Model and Application Choices}


Following the related work, we identified (1) similarity learning and (2) classification learning as the two state-of-the-art machine learning approaches selected for our comparison. 
We use both transformer-based and Recurrent Neural Network (RNN) architectures.
This design builds on prior work \cite{RackMotionPasswords2024, liebersUnderstandingUserIdentification2021a, nairUniqueIdentification500002023}, which has demonstrated that deep learning approaches are highly effective for user identification.
Previous work has shown that simpler methods, such as feature distance metrics or Multilayer Perceptron (MLP), are less effective when identifying users via motion data \cite{RackMotionPasswords2024, liebersUnderstandingUserIdentification2021a}. Both selected machine learning approaches, similarity and classification learning, allow for extensive architecture and hyperparameter optimization which is described in the upcoming section~\ref{sec:evaluation}.

\subsection{Similarity Learning (Pretrained)}

The similarity learning is a pretrainable deep learning method and is previously employed in related studies~\cite{RackMotionPasswords2024, millerUsingSiameseNeural2021, rackVersatileUser2024}.
This approach involves learning to map input data into an embedding space where distances between embeddings with the same label are minimized. 

Embeddings are multi-dimensional vectors, and distances between them can be computed using various methods.  
The model is trained so that embeddings with identical labels are close together in the embedding space, while embeddings with different labels are far apart.  
A key advantage of embeddings is that they allow identification of users who were not present during training, as each user receives a distinctive embedding based on their movements. 

\subsection{Classification (Non-Pretrained)}

The classification model is a non-pretrained deep learning method.
This method, previously employed in related studies \cite{rogersApproachUserIdentification2015, millerWithinSystemCrossSystemBehaviorBased2020, rackWhoAlyx2023, nairUniqueIdentification500002023, baldoniMovementTrafficbased2025}, has demonstrated strong performance in identifying users.
It requires that all users be present during training, as it cannot generalize to unseen users.
The model outputs a vector with one dimension per user, and the identity with the highest activation is selected as the prediction.
Although this method lacks flexibility in handling new users, it enables us to examine how well a model can handle high intra-class variance, especially when trained on motion data from diverse VR applications.

\subsection{Model Training}

The selected two classification and similarity models are trained on all VR applications (see figure~\ref{fig:all_vr_applications} and next section) to test whether they can learn consistent user representations despite differing interaction styles, motion patterns, and environmental constraints.
In contrast to the narrower evaluations of previous work, such as Baldoni et al.~\cite{baldoniMovementTrafficbased2025}, which only briefly addressed generalization and only trained the models on one application, we perform a more comprehensive analysis of model behavior in different applications.


\subsection{VR Applications}
\label{sec:vrapplications}

The main requirement for the choice of applications used to create our cross-application dataset and the model evaluation is a sufficient variety of application- and task-specific  motion patterns these applications require or involve. We selected the following five VR applications, also depicted in figure~\ref{fig:all_vr_applications}:

\textbf{1. Synth Riders} is a rhythm game where players hold two virtual balls and follow a path to dance in sync with the music. After a brief tutorial in the game, all participants play the same set of songs in the same sequence.
The game includes only a few different activities; users primarily stand and move their hands back and forth to hit the balls in the game.

\textbf{2. Superhot VR} is a first-person shooter with a unique mechanic: time only moves forward when the player moves, meaning the faster the player moves, the faster time progresses. 
All participants start at the beginning of the game and go through the integrated tutorial. 
The game includes many activities the user can perform, such as dodging enemies, shooting with a weapon, punching the enemies, or walking around.

\textbf{3. Beat Saber} is a rhythm game where players use laser swords to slash boxes in the correct direction, following the rhythm of the music. 
Participants begin with a short tutorial in the game and then play a set of songs in a predetermined order. 
This VR application has only a few different activities the user can perform: the user slashes the boxes by swinging their hands back and forth, mainly staying in place with minimal movement.

\textbf{4. Half-Life: Alyx} is a first-person shooter where players move freely around the world, solving puzzles and shooting monsters. 
Participants start later in Chapter 1 when they are given a weapon for the first time and receive a brief introduction to its use.
In this game, users have much freedom and thus can perform many different activities through various navigational options. 
They can shoot, grab objects, solve puzzles through different controller movements, and move around the room directly or via teleportation.

\textbf{5. Social VR} is represented by a diadic multi-modal conversation in a virtual environment between participants. Here, each individual interacts with an experimenter physically placed in a separate room. This Wizard-of-Oz-like setting was used to better control the flow and progress of the interaction. Conversations occur through the VR headset, which includes a built-in microphone and speakers. 
The topics are not predetermined, but most discussions center around previously played games.
These conversations are not recorded. 
In this VR application, users are free to move as they wish, with no instructions from the application. 
However, they can move around the virtual room using only physical movement; teleportation is turned off to prevent them from getting too close to each other.
During a conversation, there are many different activities that a user can perform, such as deictic gestures, beat gestures, iconic gestures \cite{kendon2004gesture, mcneill1992hand}, and many others.

Two of these—Synth Riders and Beat Saber—feature largely predefined movements, while the other two—Half-Life: Alyx and Superhot VR—involve more unconstrained motion.
Additionally, to explore identification performance in a social VR scenario, we included a fifth application: a social VR application. 
We will reveal clear distinctions in motion characteristics between the selected applications with preliminary statistical analyzes.

%% file: sections/04-datasets.tex
\section{Cross-Application Dataset Collection}
This section describes the dataset creation process involving the five VR applications.  It outlines the data collection procedure, and provides details about the participants involved. Participants were recruited through our university's participant recruitment system. 
We submitted an ethics application for the study, and the institution's ethics committee approved it.


\subsection{Procedure}
\label{sec:procedure}

\begin{figure}[tb]
 \centering 
 \includegraphics[width=0.7\columnwidth]{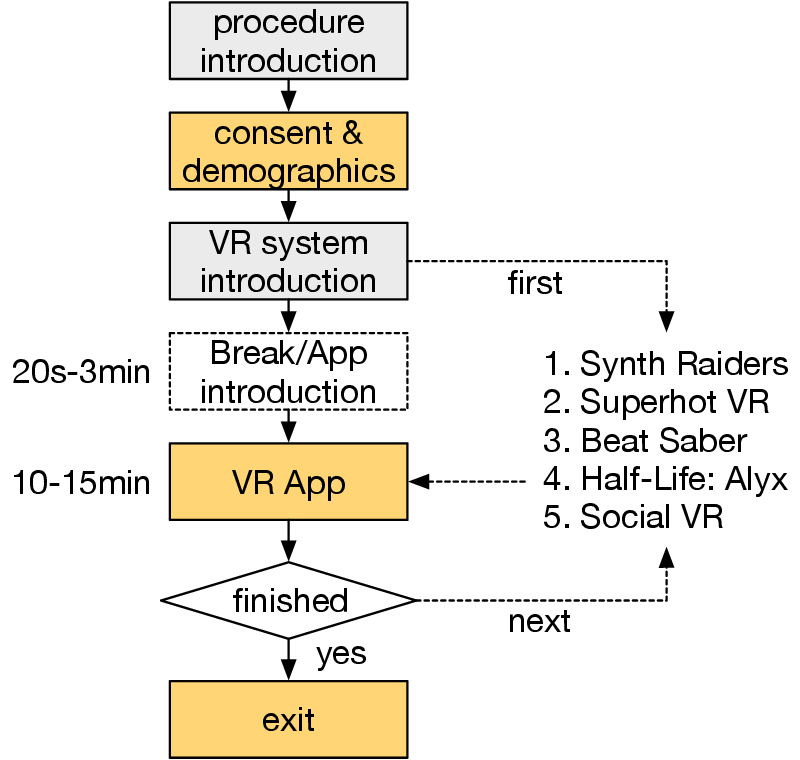}
 \caption{Procedure of the dataset collection.}
 \label{fig:data-collection}
\end{figure}

The overall procedure of the data collection process is illustrated in figure~\ref{fig:data-collection}. First participants were informed about the procedure and the various VR applications. They then consented to the data being used and published in anonymized form.
At the start, demographic data were collected for each participant, including height, weight, general VR experience, and VR games previously played. Participants were then given a brief introduction to the VR setup and, before each VR application, an introduction to the VR application and its controls. 
Each participant used the VR applications in sequence, spending 10 to 15 minutes on each, with the option to remove the headset and take short breaks between each VR application. 
The participants stood during all the VR applications.
We chose the fixed sequence of VR applications following figure~\ref{fig:data-collection} to ensure consistency and comparability of the recorded movement data between users in the respective VR applications.
The chosen sequence aimed to create an enjoyable experience by alternating between different genres, ensuring variety and minimizing the need for breaks. We selected the social VR scenario as the final application to facilitate discussion, using the previously played games as conversation starters.
 



\subsection{Recording and Implementation}
Participants used the HP Reverb G2 headset in a 10-square-meter area within our lab.
We used Python and the OpenVR~\cite{brunsPyopenvr2025} library to log the current application and continuously collect the tracking data, which includes the positions (x, y, z) and rotations (x, y, z, w) of the left and right controllers and the headset.
Additionally, we recorded the participants' field of view using OBS (Open Broadcaster Software)~\cite{OBSStudio2024}, enabling us to analyze their actions throughout the sessions.
The social VR environment was created using the Unity game engine (version 2020.3.21f1), with Photon's PUN2 architecture serving as the networking component.
All data is provided in the repository, along with further documentation detailing the individual data points.

\subsection{Sample Decription}
The dataset consists of 49 participants, 27 males and 22 females. The participants' height ranges from 161~cm to 191~cm, with an average height of 175 cm. Age ranges from 19 to 54 years, with an average age of 27. General VR experience ranges from 0 to 200 hours, with an average of 17 hours.
38 participants have never played any VR games, 7 have played one VR game, one participant has played two VR games, and three participants have played all the VR games included in the study.

%% file: sections/05-evaluation.tex
\section{Evaluation}\label{sec:evaluation}
This section describes the evaluation we applied.
Here, we detail how we preprocessed our dataset, the architecture, and the training.

\subsection{Dataset Analyzes}
\label{sec:dataset-analyses}
We analyzed the dataset using various methods to identify potential differences between the VR applications. 
This allowed us to determine whether the dataset captures a wide range of movement patterns across applications.
First, we examined user movements by calculating the average distance between the HMD and each controller in one-minute intervals across all participants.
Second, we analyzed vertical head orientation by computing the rotation angle around the X-axis based on head rotation data. 
This indicated whether participants tended to look upward or downward. 
For each participant and application, we calculated the mean and standard deviation of these angles.
We applied ANOVA and post-hoc tests to all motion and orientation metrics to assess statistical differences between applications.
For post-hoc tests, we conducted paired t-tests with Bonferroni alpha correction and all tests were conducted with $\upalpha = .05$.
We performed all analyzes in Python using the \texttt{scipy} library.

\subsection{Preprocessing}
\label{sec:preprocessing}
We preprocess the raw data used for training and evaluation to remove irrelevant information (i.e., noise), which has been shown to improve training performance \cite{rackComparisonData2022}.  
This preprocessing involves several steps based on tools developed by Rack et al. \cite{rackNavigatingKinematicMaze2024}.  
Following this, we resample the motion data to 30 FPS and convert it into a Body-Relative-Velocity (BRV) encoding.
This encoding method eliminates irrelevant information, such as the user's position or orientation within the scene, to prevent overfitting and ensure that models focus on the relevant identification signals.
To achieve this, we transform the motion sequences into a Body-Relative (BR) encoding, where each frame's positions and rotations are referenced to the local coordinate system of the HMD. 
This process also removes the HMD's position, which is consistently fixed at its local coordinate system's origin (0,0,0). 
We then calculate the first derivative between frames (BRV) to determine the positional and angular accelerations from the BR data.
After these preprocessing steps, the input sequence consists of 18 features per frame: (rot-x, rot-y, rot-z, rot-w) for the HMD and (pos-x, pos-y, pos-z, rot-x, rot-y, rot-z, rot-w) for each controller (left and right).

\subsection{Architecture and Hyperparameter Search}
\label{sec:hyperparametersearch}
\label{sec:architecture}
To optimize identification performance, we investigated various hybrid architectures combining Transformers with Gated Recurrent Units (GRUs) and Long Short-Term Memory (LSTM) networks. 
Specifically, we evaluated models where a Transformer precedes a GRU or LSTM, as well as configurations in which a GRU or LSTM is followed by a Transformer. 
Each architecture was tested with varying model sizes and layer counts for the Transformer, GRU, and LSTM components.
The architecture that achieved the best performance on our dataset was selected.  
Input sequences are first processed by a positional encoder, which adds the frame position to each motion data frame.
The data is then passed to the transformer.
The output of the transformer is followed by a GRU, which produces the final predictions by taking the values of the last hidden state.
The classification learning model includes an additional linear layer after the GRU, as the hidden size of the GRU would otherwise be too small.
The models are implemented in Python using PyTorch Lightning and PyTorch Metric Learning \cite{musgravePyTorchMetricLearning2025}.
We will publish the complete source code upon publication.

Our architecture includes a wide range of hyperparameters, each of which can significantly influence identification performance. 
As is common in machine learning, optimal hyperparameter configurations are not known a priori and must be determined through a hyperparameter search. 
We initially defined a broad search space for this purpose, as detailed in table~\ref{tab:hyperparameter}. 
We refined the search space over time by expanding or narrowing parameter ranges in response to intermediate validation outcomes and launched new sweeps accordingly. 
In total, we conducted over 1000 runs using the training and selected the configuration that achieved the highest validation accuracy. 
The code repository documentation provides a complete overview and explanation of all hyperparameters.

\subsection{Similarity-Learning Model}
\label{sec:user_identification}
For the similarity learning model, we use the architecture described in section~\ref{sec:architecture}, which uses cosine similarity between embeddings as a similarity metric.
We split our dataset of 49 users into three disjoint sets, each containing motion data from all VR applications but involving different users.  
The first 23 users were used for training, the next 9 for validation, and the final 17 for testing to evaluate overall performance.  

To evaluate the similarity learning model, we do not directly predict which user is being identified. 
To identify a user, we compare different embeddings and measure how far apart they are.
If the distance between two embeddings is small, it means they likely belong to the same user. 
If the distance is large, it’s likely a different user.
To perform this comparison, we take an embedding (called a query embedding) that we want to identify and compare it with many reference embeddings (which come from known users). 
We then check which reference embeddings are the closest to the query embedding. 
A majority voting mechanism is then used to determine the user that appears most frequently among the nearest reference embeddings, and this user is assigned to the query embedding for user identification.

From the test dataset, we calculated different embeddings for each user and each VR application based on the different movement sequences.
For different tests, we do not use all reference embeddings from all VR applications but only from certain ones in order to test how the person is identified if you only have specific movement data from them.
For example, we can calculate how accurately the model can identify a person in Synth Riders based on their gameplay data from Beat Saber.
In this case, we use the embeddings from Beat Saber as the reference embeddings and compare them to the query embeddings from Synth Riders to identify users based on the distance between the embeddings.

We use this method to compute embeddings from one VR application to serve as reference data. 
We then evaluate how accurately the model can identify individuals using query embeddings obtained from either the same or a different VR application.
To quantify performance, we compute several metrics.
First, we calculate the standard accuracy by checking whether the nearest reference embedding corresponds to the correct identity of the query embedding.
Additionally, we compute sequence-based accuracies.
For this, we consider a sequence of query embeddings obtained over time from user movement and examine which user is most frequently matched by the nearest reference embeddings across the sequence.
We then verify whether this majority identity is correct.

\subsection{Classification Learning Model}
\label{sec:methods-classification-learning}
For the classification learning model, the use of the architecture described in section~\ref{sec:architecture}.
For this, we split the dataset of 49 users such that, for each user and VR application, 45\% of the data is used for training, 20\% for validation, and 35\% for testing.

We compute several metrics to evaluate the model’s ability to identify users on the test set.
First, we calculate accuracy by measuring how well users are identified within each individual game.
Additionally, we compute a sequence accuracy metric.
Here, we consider sequences of consecutive predictions based on temporally adjacent movement data.
The final prediction is made by selecting the user most frequently predicted within the sequence.

\begin{table}[tb]
\centering
\caption{The hyperparameter space and the final hyperparameters of the Similarity Learning Model (SLM) and Classification Learning Model (CLM).}
\label{tab:hyperparameter}
\small
\begin{tabular}{l|l|l|l|l}
\hline
& Hyperparameter &  Search Space  & SLM & CLM \\
\hline
  General & embedding size & [128, 1024] & 480  & - \\
          & learning rate & [0.001, 1e-05] & 0.00098 & 0.0004 \\
          & dropout frames & [0, 0.5] & 0.3 & 0.3\\
          & dropout global & [0, 0.5] & 0.2 & 0.2\\
          & frame step size & [30, 150] & 50 & 100\\
         & windows size & \{150, 300, 450, 600\} & 450 & 600\\
  Transformer & d\_model & [128, 1024]  & 320 & 704 \\
              & number of layers & [1, 3] & 1 & 2 \\
              & feedforward dim & [128, 1024] & 960 & 640 \\
              & nhead & \{4, 8, 16\} & 16 & 8  \\
  GRU         & hidden size &  [128, 512] & 480 & 512 \\
              & layers & [1,2] & 2 & 1\\
              & dropout & [0, 0.2] & 0.1 & 0.1 \\
  
\hline
  \end{tabular}%
\end{table}

%% file: sections/06-results.tex
\section{Results}
\label{sec:results}
In this chapter, we present the results of our data analysis and the outcomes of the machine learning models.
For the similarity learning model in particular, we conducted several evaluation tests, as this type of model allows performance to be assessed in various ways using a single trained instance, as described in section~\ref{sec:user_identification}.
In contrast, the classification learning model only allows us to test how well users can be identified within the same VR application.
It does not support evaluating cross-application accuracy without training a separate model for each application.

\subsection{Dataset Analyzes}

\begin{table}[h!]
\centering
\caption{Pitch of HMD and the average movement distance (in m) for HMD and controllers (Con. 1: left controller, Con. 2: right controller) for each VR application.}
\label{tab:movement_distances}
\begin{tabular}{l|l|l|l|l|l}
\hline
 & \multicolumn{3}{c|}{\textbf{Movement Distance}} & \multicolumn{2}{c}{\textbf{Pitch}} \\ \hline
\textbf{Application} & \textbf{HMD} & \textbf{Con. 1} & \textbf{Con. 2} & \textbf{Mean} & \textbf{Std} \\ \hline
Synth Riders   & 6.61           & 30.08               & 30.81    & 1.13    & 4,71    \\ 
Superhot VR    & 12.81          & 23.65               & 33.44    & 9.11   & 4.97      \\ 
Beat Saber     & 5.86           & 42.12               & 39.19    & 6.68   & 4.90     \\
Half-Life: Alyx & 5.73           & 14.87               & 12.21    & 17.24  & 6.40   \\ 
Social VR Scenario     & 3.34           & 7.32                & 8.13     & 2.69   & 5.70        \\ 
\hline
\end{tabular}
\end{table}
Table~\ref{tab:movement_distances} summarizes the average movement distances and head pitch angles across the five VR applications.
Head movement distances are highest in Superhot VR, approximately twice as large as those in Beat Saber, Synth Riders, and Half-Life: Alyx, and significantly higher than in social VR application.
Controller movement distances are most pronounced in Beat Saber, followed by Synth Riders, Superhot VR, and Half-Life: Alyx, with the lowest values observed in social VR application.
A one-way repeated measures ANOVA indicated significant differences in movement distances between applications:
head movement distance: $F(4, 44) = 123.32$, $p < .001$;
left controller movement distance: $F(4, 44) = 178.83$, $p < .001$;
right controller movement distance: $F(4, 44) = 176.94$, $p < .001$.
Post-hoc tests indicated significant differences in head and controller movement distances between all applications, except for head movements between Beat Saber, Synth Riders, and Half-Life: Alyx, and for right controller movements between Superhot VR and Synth Riders.

Regarding head pitch, the average pitch angle remained close to neutral in Synth Riders, Beat Saber, and social VR application, indicating a level gaze.
In contrast, participants tended to look upward more in Superhot VR, with the highest mean pitch angle observed in Half-Life: Alyx.
The standard deviations of head pitch were relatively consistent across applications, ranging from 4.71 to 6.40 degrees.
A one-way repeated measures ANOVA indicated significant differences in head pitch angles across applications, $F(4, 45) = 68.55$, $p < .001$.
Post-hoc tests indicated that all differences were significant, except between Superhot VR and Beat Saber, as well as between Synth Riders and the social VR application.
Due to space constraints, detailed post-hoc test statistics are provided in the supplementary material available via Google Drive.

\begin{figure}[tb]
 \centering 
 \includegraphics[width=\columnwidth]{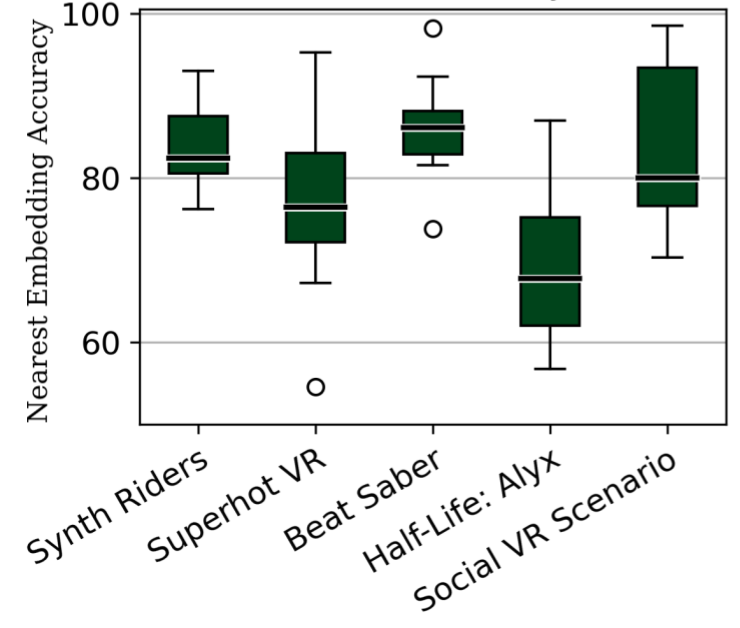}
 \caption{This figure shows the nearest embedding accuracy of user identification across VR applications using the similarity learning model, with all applications as reference data. The box plots display user-wise accuracy distributions for each game. A black line marks the median. 
}
 \label{fig:general_results}
\end{figure}

\begin{figure*}
 \centering 
 \includegraphics[width=\textwidth]{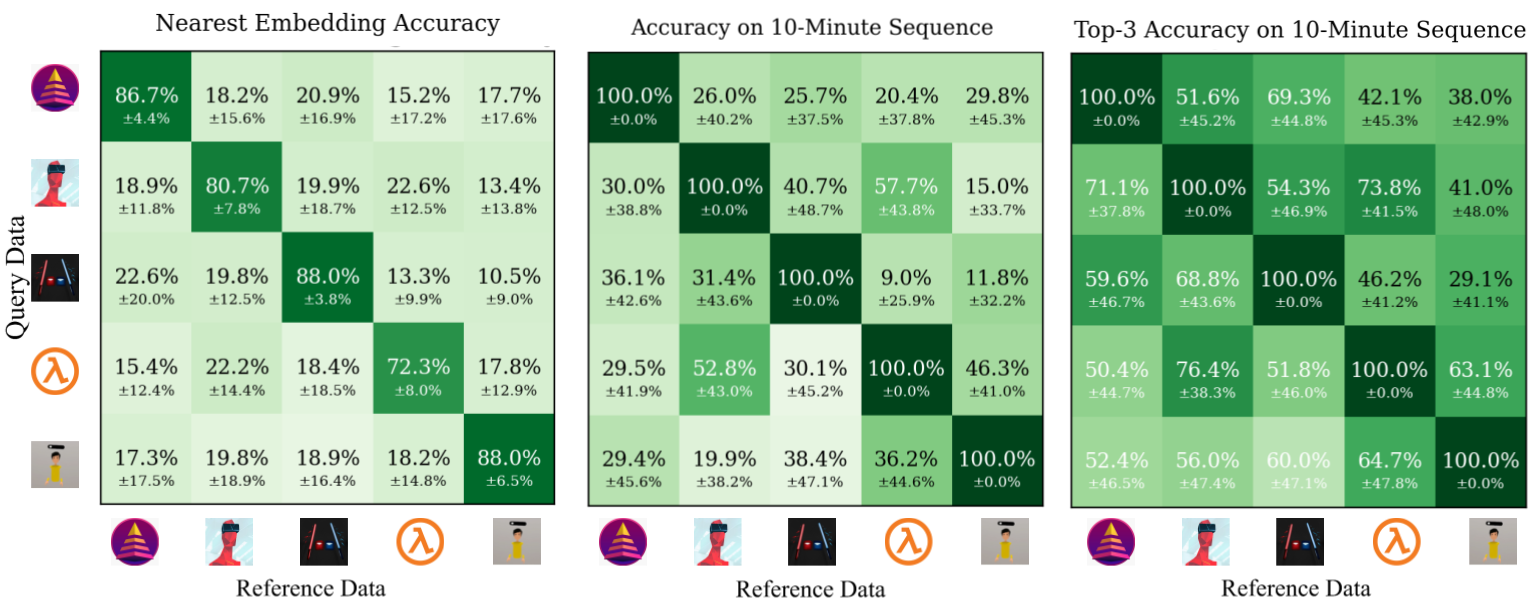}
 \caption{These figures show heatmaps for different metrics, using one application as reference and another as query. The games are listed top to bottom: Synth Riders, Superhot VR, Beat Saber, Half-Life: Alyx, and Social VR Scenario.
}
 \label{fig:multiple_small_heatmap}
\end{figure*}

\subsection{Similarity Learning}
As described in section~\ref{sec:methods-classification-learning}, we trained and evaluated a pretrainable similarity learning model using the dataset.
\label{sec:similarity-learning}

\subsubsection{Overall Performance}
To evaluate the model's overall performance in user identification, we used data from all users and VR applications as a reference.
This approach aimed to assess the general capability of the similarity learning model.
We compute the accuracy directly from the nearest embeddings and do not yet consider a larger time window of the user's interactions, as is done with sequence accuracy.
On average, the model achieved an accuracy of 78.5\% across all users and VR applications.
Figure~\ref{fig:general_results} shows the mean accuracy for each VR application, computed using only query embeddings from the respective application.
The results indicate minor variations in accuracy between applications.
Specifically, user identification accuracy ranged from 67,7\% in Synth Riders to 86,0\% in Beat Saber.

\subsubsection{Different Application in Reference and Query}
\label{sec:cross-application-identification}
In section~\ref{sec:user_identification}, we present results when both references and queries are taken from either the same or different applications.
In this step, we identify the user whose reference embedding is closest to the query embedding.
When references and queries are taken from the same application (represented by the diagonal on the left heatmap in figure~\ref{fig:multiple_small_heatmap}), user identification accuracy ranges from 72.3\% to 88.0\%, with an average of 83.1\% and a mean standard deviation of 6.1\%.
In contrast, when references and queries are taken from different applications (all off-diagonal entries on the left heatmap in figure~\ref{fig:multiple_small_heatmap}), accuracy drops to an average of 18.0\%, with a range between 10.5\% and 22.6\%.
The corresponding standard deviations range between 9.0\% and 20.0\%, with an average of 15.1\%.

\subsubsection{Sequence Accuracy}
As described in section~\ref{sec:user_identification}, we compute the sequence accuracy.
As shown in the middle heatmap in figure~\ref{fig:multiple_small_heatmap}, accuracy increases significantly when the user is observed for 10~minutes.
When references and queries originate from the same application (represented by the diagonal), accuracy reaches 100\%.
When references and queries stem from different applications (i.e., all elements off the diagonal), accuracy ranges from 9.0\% to 57.7\%, with an average of 30.8\%.
The heatmap shows that when both the reference and query come from the applications Superhot VR and Half-Life: Alyx, accuracy exceeds 50\%.

\subsubsection{Top-3 User Identification}
\label{sec:top-3-user-identifications}
Since cross-application accuracy for identifying the correct user is relatively low, we investigated whether our model could at least narrow down the prediction to the top three candidates.
The majority voting approach we used not only predicts the most likely user but also returns a ranked list of the top candidates.
Rather than focusing solely on the top-1 prediction, we evaluated whether the correct user appeared among the top three candidates with the highest number of similar reference embeddings.
The right heatmap in figure~\ref{fig:multiple_small_heatmap} shows that the 10~minutes sequence accuracy reaches 100\% when both the reference and query samples originate from the same application (i.e., along the diagonal), as measured by whether the correct user appears among the top three candidates.
When reference and query samples come from different applications (i.e., the off-diagonal elements), accuracy ranges between 29.1\% and 76.4\%, with an average of 56.0\%.

\subsection{Classification Learning}
\label{sec:results-classification-learning}
As described in section~\ref{sec:methods-classification-learning}, we trained and test a non-pretrainable classification model using the dataset.
The model achieved a maximum validation accuracy of 46.7\%, indicating its capability to identify individuals correctly.
Using this final model, we assessed its generalization performance on the test data, achieving an accuracy of 43.2\%.
To further evaluate the model, we computed the sequence accuracy over a 2:30~minutes window.
Due to the separation of training, validation, and test sets, the number of contiguous sequences was limited, resulting in a sequence accuracy of 46.2\%.
We also evaluated user identification performance across different applications by separately measuring overall accuracy and 2:30~minutes sequence accuracy on the test dataset.
The model achieved 46.5\% accuracy and 49.6\% sequence accuracy on Synth Riders, 41.9\% accuracy and 45.3\% sequence accuracy on Superhot VR, 68.7\% accuracy and 73.0\% sequence accuracy on Beat Saber, 27.8\% accuracy and 30.7\% sequence accuracy on Half-Life: Alyx, and 28.4\% accuracy and 29.4\% sequence accuracy in the social VR application.

%% file: sections/07-discussion.tex
\section{Discussion}
In this work, we investigated the capabilities of two prominent state-of-the-art machine learning models to identify individuals based on their motion patterns across different XR applications. 

Our results suggest that the ability of recent models, whether pretrainable or non-pretrainable, is yet limited when identifying individuals across different applications. 
However, our dataset includes a wide range of VR applications with specific and unspecific movements, thus encompassing many different activities.
This diversity provides a valuable opportunity for researchers to train and evaluate new machine learning models.
It enables a detailed analysis of how future machine learning approaches might perform within individual VR applications and across multiple applications.

\subsection{Within-App Identification}
As shown in prior work, users can be reliably identified within the same application \cite{rackVersatileUser2024, rogersApproachUserIdentification2015, millerWithinSystemCrossSystemBehaviorBased2020, rackWhoAlyx2023, nairUniqueIdentification500002023}.
Our similarity learning model confirms this finding: although it was trained on all data from the application, it can accurately identify new users when prior data from the same application is available.
Both rhythm games yielded high identification scores, likely due to minimal activity variation, as users usually performed similar motions.
Identification accuracy in Superhot VR was comparable, which is reasonable since user activities, such as moving or shooting, remained relatively consistent.
Accuracy in the social VR application was also comparable, as the range of movements in this setting is not overly complex.
In contrast, Half-Life: Alyx presents a broader range of possible activities, making the identification task more challenging for the model.
Figure~\ref{fig:general_results} initially suggests that the similarity learning model performs worse than previous approaches.
This is likely due, on the one hand, to a higher variance in the data, and on the other hand, to the fact that we are directly evaluating the output—i.e., the nearest embedding.
However, as shown in the middle heatmap in figure~\ref{fig:multiple_small_heatmap}, when a longer sequence is considered, the accuracy increases significantly, reaching 100\%.

\subsection{Cross-App Identification}
Our similarity learning model demonstrates relatively strong performance when identifying users within a single application.  
However, cross-application identification—recognizing a user across different contexts—proves significantly more challenging.
The average accuracy in this setting is just 18.0\%, which is much better than random guessing, which is 5.8\% ($100 - \text{user\_number} = 100 / 17 = 5.8\%$).
This is also evident in our classification learning model.
Similar to the similarity learning model, it was trained using all applications, but unlike the similarity learning model, it included all users during training.
Despite this, the classification learning model still struggles to identify individual users, achieving only 43.5\% accuracy.
Baldoni et al.~\cite{baldoniMovementTrafficbased2025} reported similar results with their classification learning model when they trained it on one application and then attempted to identify users in a second application.
When we look at the individual accuracy scores from the VR applications, we see that the model has fewer difficulties identifying users in Beat Saber compared to the other applications.
This might be because the user movements in Beat Saber were the most consistently distinguishable for the classification learning model. 
This may have allowed the model to focus on these patterns during training when it had access to data from all VR applications.

A likely reason for the low accuracy of both models is the pronounced difference in movement patterns across applications.
Our statistical analyzes of the dataset also confirm these differences, revealing clear distinctions in two dimensions across the datasets.
In applications such as Beat Saber or Synth Riders, users exhibit more intense movements---as measured by controller travel distance---compared to social VR application or Superhot VR.
Head movements also differ significantly.
This variability makes it difficult for the models to learn a consistent representation of user identity across contexts.

The model's accuracy can be improved by calculating the sequence accuracy and observing the user over a longer period.
For example, in the similarity learning model, accuracy increases significantly—exceeding 50\%—when observing users across applications such as Half-Life: Alyx and Superhot VR.
This improvement is likely because both VR applications involve more undefined or exploratory movements, leading to a greater overlap in motion patterns and, consequently, more similar embeddings.
This also highlights an advantage of the similarity learning approach, in addition to the fact that new users can be identified after training.  
Since movements are translated into embeddings rather than directly mapped to users, the model can assign similar embeddings to similar movements, even when originating from different applications.  
As a result, it can still correctly identify the user, despite potential difficulties distinguishing between the contexts.
A similar, albeit less pronounced, effect is observed in the classification learning model: accuracy also increases—by over 5\%—when the user is observed for a longer duration.
Compared to the similarity learning model, the smaller gain is likely due to the shorter observation window: only 2:30~minutes for the classification model versus 10~minutes for similarity learning.
This discrepancy stems from how the dataset was partitioned for training, validation, and testing.
When analyzing the top-3 accuracy of the similarity learning model, user identification improves further, indicating that the model indeed captures meaningful user-specific patterns.

The core challenge is bridging the gap between the different types of movement exhibited by the same user across different applications.  
These intra-user differences are often greater than inter-user differences within similar applications, further hampering generalization.  
Increasing architectural complexity—through additional layers or alternative layer types—does not appear to be a sufficient solution.  
Instead, a fundamental rethinking of modeling approaches may be required.
Despite these challenges, the potential for improved cross-application identification is evident, as shown by the top-3 prediction results from the similarity learning model.  
Still, it remains an open question whether an ideal model can truly transfer user-specific movement profiles from one application context to another.  
At present, the threat of cross-app identification can still be considered moderate.  
Nonetheless, our dataset—with its broad spectrum of movement styles, from highly structured to entirely unstructured—provides a valuable foundation for future research into this question.

\section{Future Work}
The state-of-the-art models used were developed with previous datasets from individual VR applications. 
Enhancing these models could improve their reliability and accuracy through several avenues of future work.
One approach involves refining the model architectures by incorporating strategies from other areas of machine learning and adapting them to user identification. 
Additionally, improvements in data preprocessing could further enhance model performance.
Currently, our models rely primarily on motion data for user identification. 
Integrating additional contextual information, such as details about the XR application or types of movements performed, could increase user identification accuracy, potentially improving cross-application performance.

When considering the use of additional datasets for user identification across various applications, it is crucial to address privacy concerns. 
In future work, the dataset can also be used to test current approaches for anonymizing motion data and potentially improve these methods.

\section{Limitations}
Our work exhibits the following limitations that should be considered when interpreting the results.
Firstly, our dataset narrow range of ethnicities may not fully capture the diversity of motion profiles across different cultures and populations.
Including a more diverse participant pool would provide a more comprehensive understanding of these variations.
Secondly, while we present the dataset and initial results, these findings were obtained using current models. 
They should not be considered definitive when determining whether individuals can be reliably identified across VR applications.

\section{Conclusion}
This paper explored the capacity of state-of-the-art similarity learning and classification learning model to identify users based on their motion data across different VR applications.
For that, we developed and released a novel dataset comprising motion data from 49 users across five distinct VR applications, allowing us to assess identification performance both within and across applications.
Our results demonstrate that while models can achieve high accuracy in identifying users within a single application, cross-application identification remains significantly more challenging.
While cross-application identification is not yet highly accurate, it already outperforms random chance, raising privacy concerns regarding future advancements.
As motion-based identification models continue to evolve, the provided dataset will be valuable to benchmark new models and assess the growing risk of identification in XR environments.
